**PREPRINT**

# Hall effect in the normal state of high Tc cuprates

J. Bok[a] and J. Bouvier[a]

[a]Solid State Physics Laboratory - ESPCI. 10, rue Vauquelin. 75231 PARIS Cedex 05, FRANCE.

**Abstract**

We propose a model for explaining the dependence in temperature of the Hall effect of high Tc cuprates in the normal state, in various materials (LSCO, YBCO, BSCCO, GdBCO). They all show common features: a decrease of the Hall coefficient $R_H$ with temperature and a universal law, when plotting $R_H(T)/R_H(T_0)$ versus $T/T_0$ where $T_0$ is defined from experimental results. This behaviour is explained by using the well known electronic band structure of a $CuO_2$ plane, showing saddle points at the energies $E_S$ in the directions $[0, \pm \pi]$ and $[\pm \pi, 0]$. This is well confirmed by photoemission experiments. We remark that in a magnetic field, for energies $E > E_S$ the carrier orbits are hole-like and for $E < E_S$ they are electron-like, giving opposite contributions to $R_H$. We are able to fit all experimental results for a wide range of hole doping ($p_{h0}$) ($0.09 < p_{h0} < 0.30$), and to fit the universal curve. For us $k_BT_0$ is simply $E_F - E_S$, where $E_F$ is the Fermi level varying with the doping.

*PACS:* 74.72.-h; 74.25.Fy; 73.50.Jt
*Keywords:* High Tc cuprates, Hall effect, normal state

## 1. Introduction

Many measurements of the Hall effect in various high Tc cuprates have been published [1-5]. The main results are the following :

(i) at low temperature T, $R_H \approx 1/p_{h0}e$, where $p_{h0}$ is the hole doping, when T increases $R_H$ decreases, and for highly overdoped samples becomes even negative [1]

(ii) these authors are also able to define a temperature $T_0$ , where $R_H$ changes its temperature behaviour, and such $R_H(T)/R_H(T_0)$ versus $T/T_0$ is a universal curve for a large doping domain (from $p_{h0} = 0.10$ to $p_{h0} = 0.27$).

We show that we can explain these results by using the band structure for carriers in the $CuO_2$ planes. In particular, the existence of hole-like and electron-like constant energy curves, which give contributions of opposite sign to the Hall coefficient $R_H$. The transport properties explore a range of energy $k_BT$ around the Fermi level, when T is increased more and more carriers are on the electron like orbits, resulting in a decrease of $R_H$.

## 2. Calculation of the Hall coefficient

The constant energy surfaces of carriers in the $CuO_2$ planes are well describe by the following formula:

$$E_k = -2t(\cos X + \cos Y) + 4t' \cos X \cos Y + E_F - E_S + 4t' \quad (1)$$

where t is the transfer integral between the first nearest neighbours, t' between the second nearest neighbours, a the lattice parameter, $E_S$ the energy of the saddle point (Van Hove singularity, VHS) and $E_F$ is the Fermi level, which varies with the doping [6]. These electronic structures have been intensively studied by angular resolved photoemission (ARPES) in BSCCO [7], and more recently by Ino et al [8] in LSCO, for a wide range of Strontium doping. From ARPES, experimental values for t and t' are obtained. It is very clearly seen [8] that the Fermi level crosses the saddle points, at $E_S$, (VHS) for a hole doping of $p_{h0} \approx 0.22$. For $E > E_S$ the orbits are hole-like, and for $E < E_S$ they are electron-like.

* To whom correspondence should be addressed:
E-mail : julien.bok@espci.fr

To compute the Hall coefficient we use the formula obtained by solving the Boltzmann equation. In the limit of low magnetic fields B, perpendicular to the $CuO_2$ plane, $\mu B << 1$, where $\mu$ is an average mobility of the carriers, $R_H$ is given by:

$$R_H = \frac{\sigma_{xy}}{\sigma_{xx}^2} \frac{1}{B} \tag{2}$$

where $\sigma_{xy}$ and $\sigma_{xx}$ are the components of the conductivity tensor. We follow the approach given by N. P. Ong [9] :

$$\sigma_{xy} = \int_{E_{\min}}^{E_{\max}} \left( -\frac{\partial f_0}{\partial E} \right) \sigma_{xy}(E)\, dE \tag{3}$$

where $f_0$ is the Fermi Dirac distribution function, $E_{min}$ and $E_{max}$ are the bottom and the top of the band, and $\sigma_{xy}(E)$ is $\sigma_{xy}$ computed on a constant energy surface.

For metals, where $k_B T << E_F$, $\sigma_{xy}$ is usually chosen as $\sigma_{xy} = \sigma_{xy}(E_F)$, computed on the Fermi surface only, this is done by N. P. Ong [9]. In our case, $k_B T$ is not small compared to $E_F - E_S$, so when T increases the electron-like orbits as well as the hole-like orbits are populated. The electron-like orbits give a negative contribution to $R_H$, so that $R_H$ decreases with temperature. This is our original approach to the problem. To compute $R_H$, we use the following method: we compute first $\sigma_{xy}(E)$ using the Ong approach. The idea is to draw the $\vec{l}$ curve swept by the vector $\vec{l} = \vec{v}_k\, \tau_k$ as $\vec{k}$ moves around the constant energy curve (C.E.C.). Then $\sigma_{xy}$ reduces to:

$$\sigma_{xy} = \frac{2e^3}{\hbar^2} A_l\, B \tag{4}$$

where $A_l$ is the area enclosed by C.E.C., in the $(l_x, l_y)$ plane. There may be secondary loops in the $\vec{l}$ curve. When the C.E.C. is non-convex, the $\vec{l}$ curve presents several parts where the circulation are opposite (see Ref 9 Fig. 2). Then the effective density of carriers that must be taken in computing $\sigma_{xy}$ is $n_e' = \Gamma\, n_e$ for the electron-like orbits, with $\Gamma < 1$, and $p_h' = p_h$ for the hole-like orbits, because for the hole-like orbits we can see that the C.E.C. have no non-convex parts. Finally we obtain for the Hall coefficient:

$$R_H = \frac{V}{e} \frac{p_h - b^2\, n_e'}{\left(p_h + b\left(p_{h0} - p_h\right)\right)^2} \tag{5}$$

where $b$ is the ratio of the average mobilities of the carriers on the electron and hole like orbits. That is the mean value of $\langle \tau / m \rangle$, where $\tau$ is the relaxation time and m the effective mass. V is the volume of the unit cell. We adjust the Fermi level so that the total number of carriers $p_{h0}$ remains constant. To compute $\Gamma$, we must know the scattering mechanisms and evaluate $\Gamma$. $\Gamma$ was computed by Ong [9] assuming a constant $\vec{l}$, but this is not valid in our case because $\vec{l}$ is very small near the saddle points (hot spot), both $v_k$ and mainly the relaxation time $\tau_k$ are small at this point. So we estimated a much smaller value of $\Gamma$, around $\Gamma = 0.2$ for E near $E_S$ and going to $\Gamma = 1$ when E approaches $E_{min}$. We choose a function $\Gamma(E)$, varying from $\Gamma(E_S)$ to 1 for $\Gamma(E_{min})$.

$$\Gamma(E) = 1 - \alpha\left(E - E_{\min}\right)^n$$

$$\alpha = \frac{1 - \Gamma(E_S)}{\left(E_S - E_{\min}\right)^n} \quad , \quad with \quad n = 1$$

$n_e^{'}$ , $p_h$ are given by the following formulae:

$$n_e^{'} = \int_{E\min}^{Es} A_e(E)\ \Gamma(E)\left(-\frac{\partial f_0}{\partial E}\right) dE \qquad (6)$$

$$p_h = \int_{Es}^{E\max} A_h(e)\ \Gamma(E)\left(-\frac{\partial f_0}{\partial E}\right) dE \qquad (7)$$

$A_e = A_l$ is the area enclosed by the electron-like surfaces for $E < E_S$, $A_h = A_l$ is the area enclosed by the hole-like surfaces for $E > E_S$. $E_{max}$ is determined in order to only take into account the holes added to the lower half-band. So we obtain for T = 0 K the number of free hole carriers $p_{h0}$ for the Hall number $n_H = V/(R_H e) = p_{h0}$. The scattering mechanism being probably the same for the electron and the hole orbits, which are very similar along the (1,1,0) direction, then we assume b = 1.

## 3. Results

The results of our calculations and their comparison with the experimental results are given in Fig. 1, 2, 3, 4.

When the authors of the experimental results give only the concentration x of doping atoms, and the critical temperature Tc we have to evaluate the actual hole doping $p_{h0}$ using the universal phase diagram of Tc versus hole doping for high Tc superconductors [10].

For the theoretical results of figures 1, 2, 3 we use the following parameters : t = 0.18 eV, t' = 0.04328 eV, 2t'/t = 0.48, $\Gamma(E_S) = 0.2$. These values of t and t' means that the shape of the Fermi surfaces changes when we cross the critical doping $p_{h0} \approx 0.22$. This is also seen in the photoemission curves reported by Ino et al [8].

In Figure 2, we can see the representation of the universal law $R_H(T)/R_H(T_0)$ versus $T/T_0$, where $T_0$ is defined experimentally by the fact that $R_H$ becomes almost constant above this temperature [1-4]. In our model this temperature is given by $2k_BT_0 = E_F - E_S$, this shows that this universal behaviour is due to the 2D band structure, in which the shift $E_F - E_S$ is connected to the hole doping. This is very natural in our approach, because the factor $(E_F - E_S)/k_BT$ enters the Fermi-Dirac distribution.

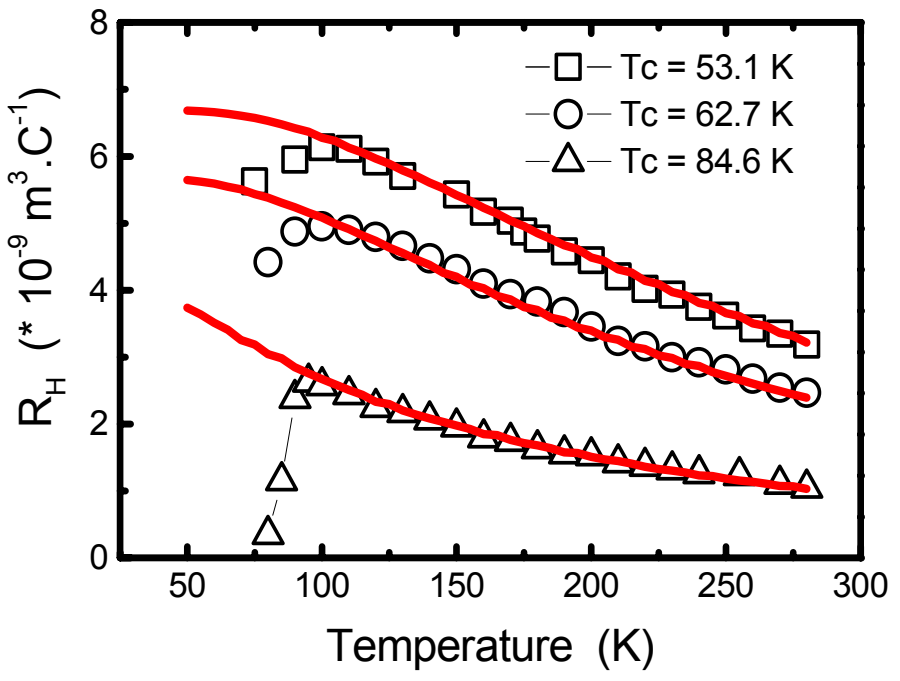

Figure 1 –

symbols: experimental $R_H(T)$ given by Matthey et al (Ref. 4) in $GdBa_2Cu_3O_{7-\delta}$

full lines: theoretical fits

theoretical hole doping level $p_{h0} = 0.10$ for the experimental Tc = 53.1 K

theoretical hole doping level $p_{h0} = 0.12$ for the experimental Tc = 62.7 K

theoretical hole doping level $p_{h0} = 0.16$ for the experimental Tc = 84.6 K

The calculations are made with : t = 0.18 eV , t' = 0.04328 eV , 2t'/t = 0.48, $\Gamma(E_S) = 0.2$

a)

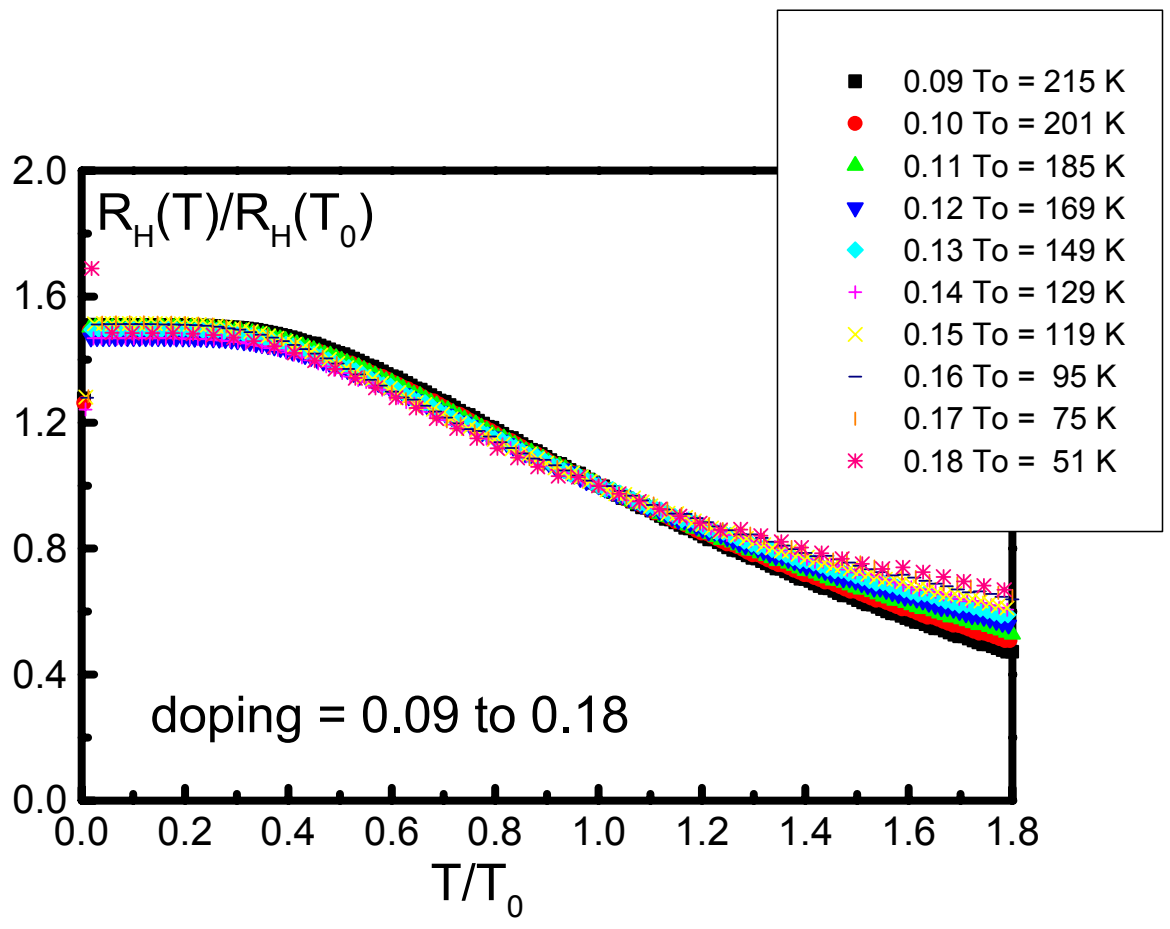

b)

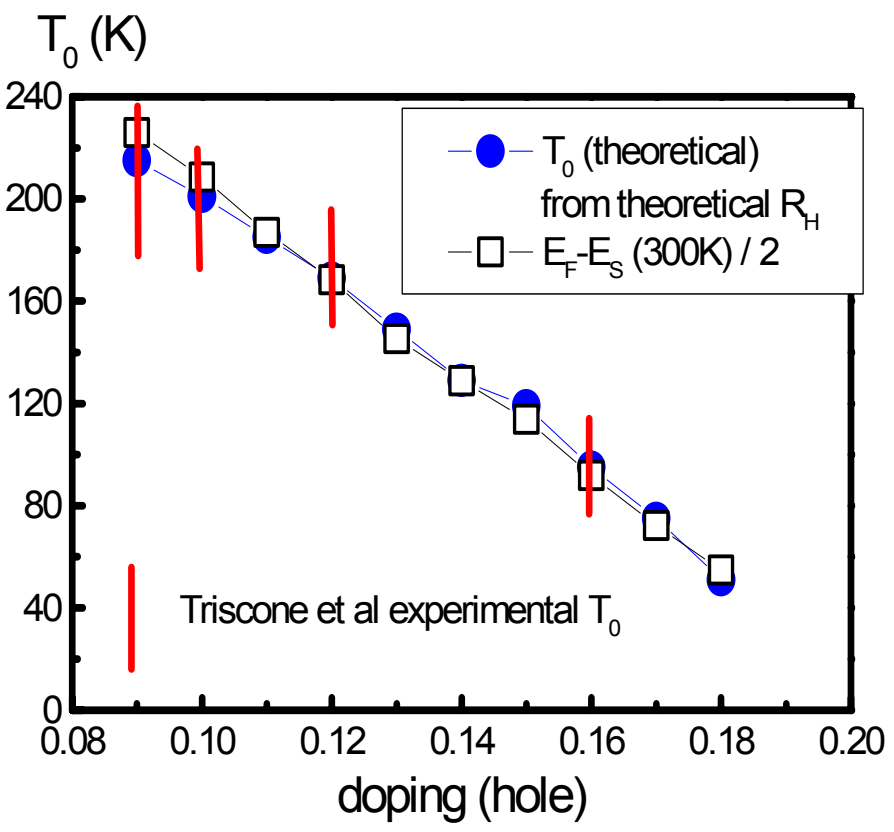

Figure 2 –

a – universal law $R_H(T)/R_H(T_0)$ versus $T/T_0$ for various hole doping levels, from 0.09 to 0.18.

b – calculated $T_0$, $2k_BT_0 = E_F - E_S$, compared with the experimental $T_0$ given by Matthey et al (Ref. 4)

We see that the agreement of our fits with the experiments are excellent. There is a small discrepancy between the values of our theoretical $R_H$ and the experimental values. We think that this is due to the inhomogeneities in the material and to the way to carry out the $R_H$ measurements. This can may be explained by the evaluation of the experimental volume V. We use in our calculation the unit cell volumes:

$V_{LSCO} \cong 189\ 10^{-30}$ m$^3$ for LSCO

and $V_{YBCO} \cong 174\ 10^{-30}$ m$^3$ for YBCO.

The experimental value of $R_H$ is determined by the geometrical aspect of the sample (the thickness in particular). This value is evaluated assuming that the current flow is homogeneous throughout the sample, this is not always true. We find a discrepancy between 1.5 and 2 in the case of YBCO [2] and GdBCO [4], a larger discrepancy is found in the case of LSCO [1]. In this later case, the authors find different $R_H$ results for the same doping, with various compounds (single crystals and thin films).

Anyway, adjusting our values for $R_H$, at low temperature, we can fit many experimental results, for the three different compounds. We also use a rigid band model, where the bandwidth does not change with the doping. This is not exactly the case as shown in the photoemission experiments [8], but the effect is small and does not change our conclusions.

From overdoped to lightly underdoped samples the upturns (in $n_H$) or downturns (in $R_H$), at low temperature, in the experimental curves are due to the occurrence of the superconductivity transition.

a)

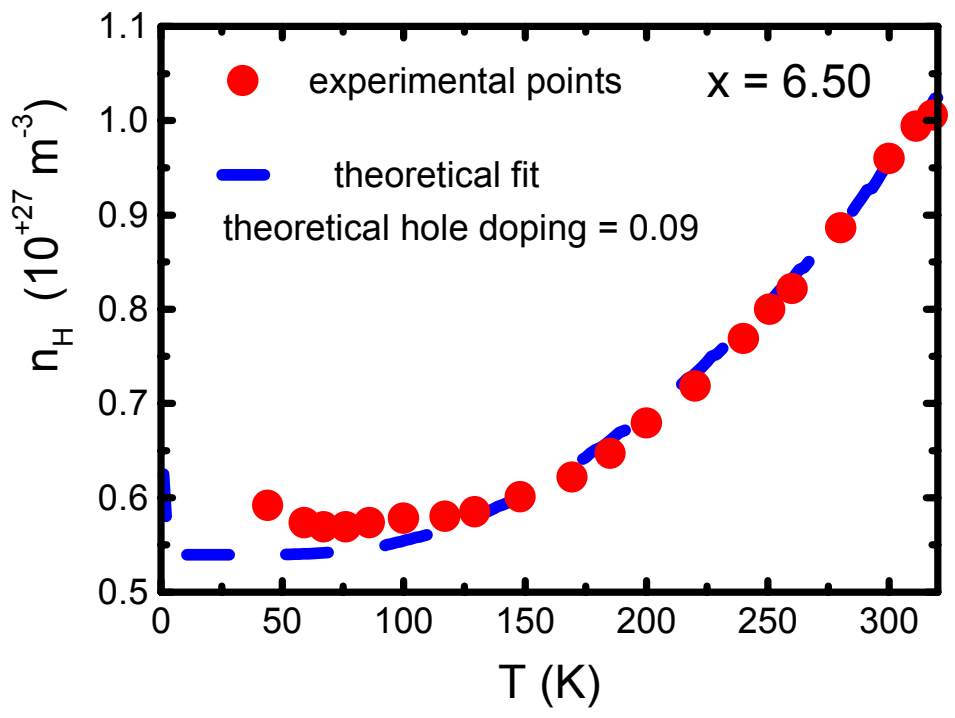

b)

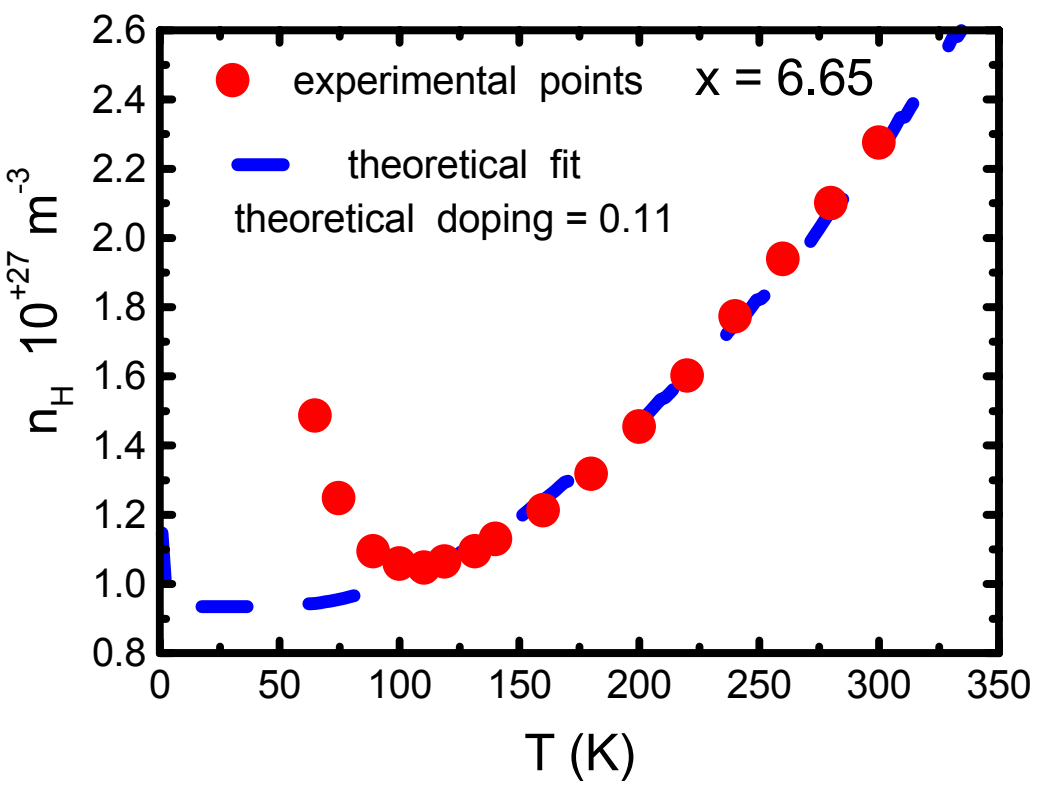

c)

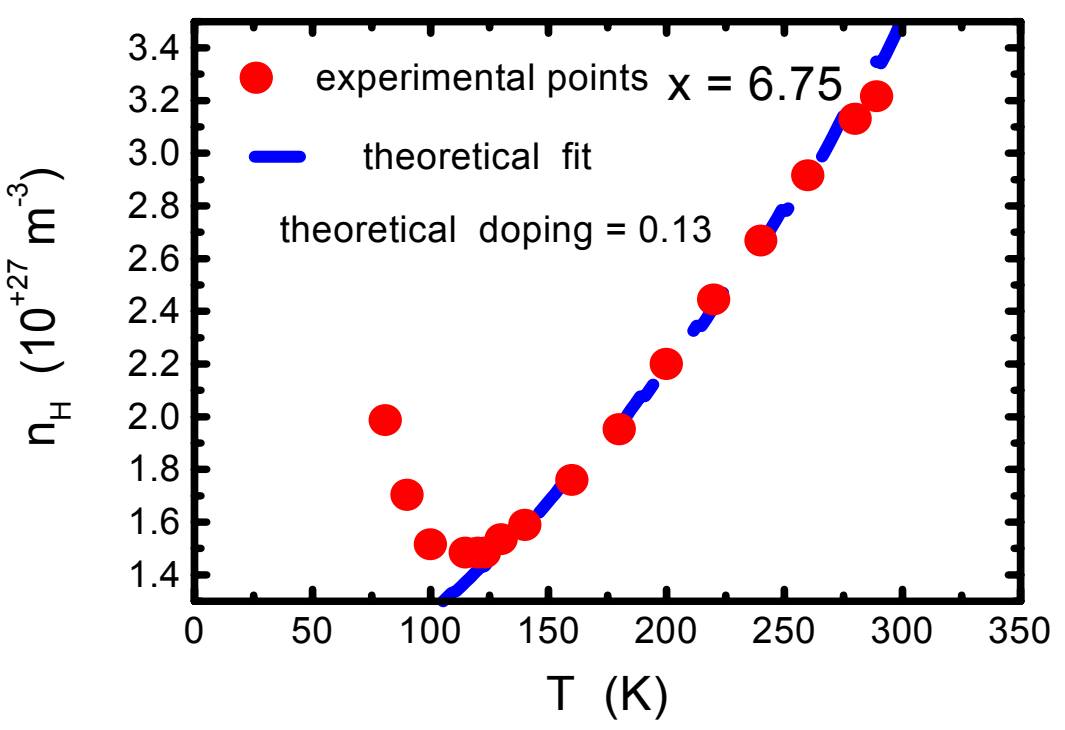

d)

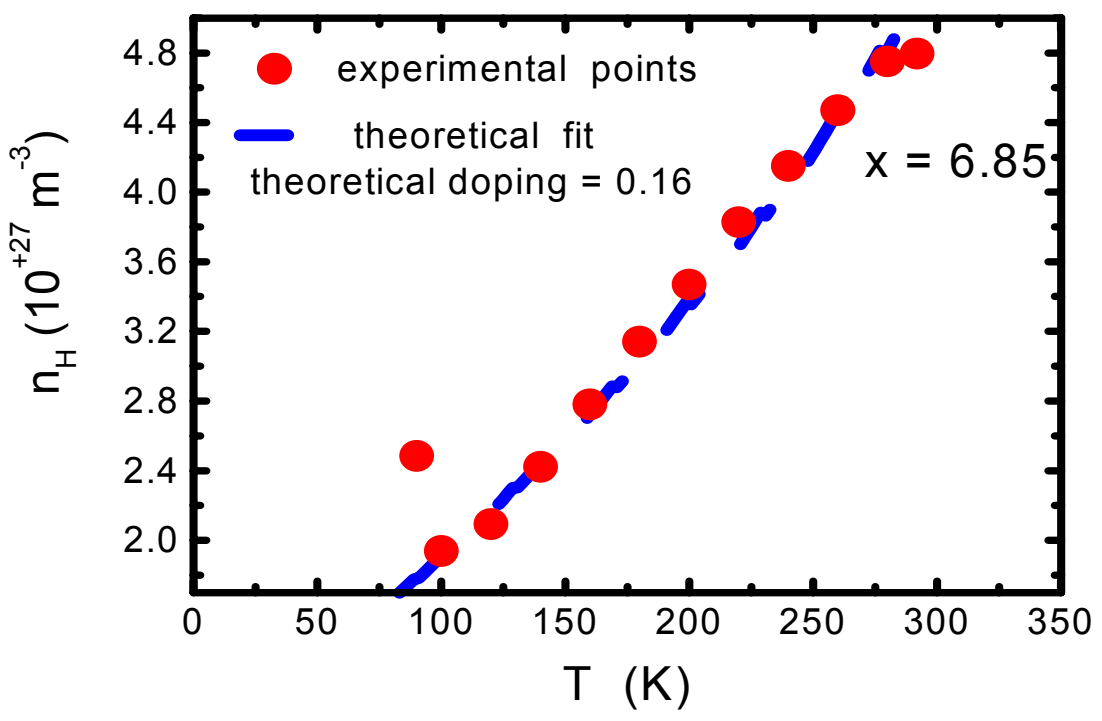

Figure 3 -

Filled circles: experimental $n_H(T) = V/(R_H e)$ given by Wuyts et al (Ref. 2) in $YBa_2Cu_3O_x$.

Dashed lines: theoretical fits

a) x=6.50, theoretical hole level = 0.09
b) x=6.65, theoretical hole level = 0.11
c) x=6.75, theoretical hole level = 0.13
d) x=6.85, theoretical hole level = 0.16

The calculations are made with: t = 0.18 eV, t' = 0.04328 eV, 2t'/t = 0.48, $\Gamma(E_S) = 0.2$
then we obtain the same universal law as in Figure 2a, expressed in $n_H(T)/n_H(T_0)$ .

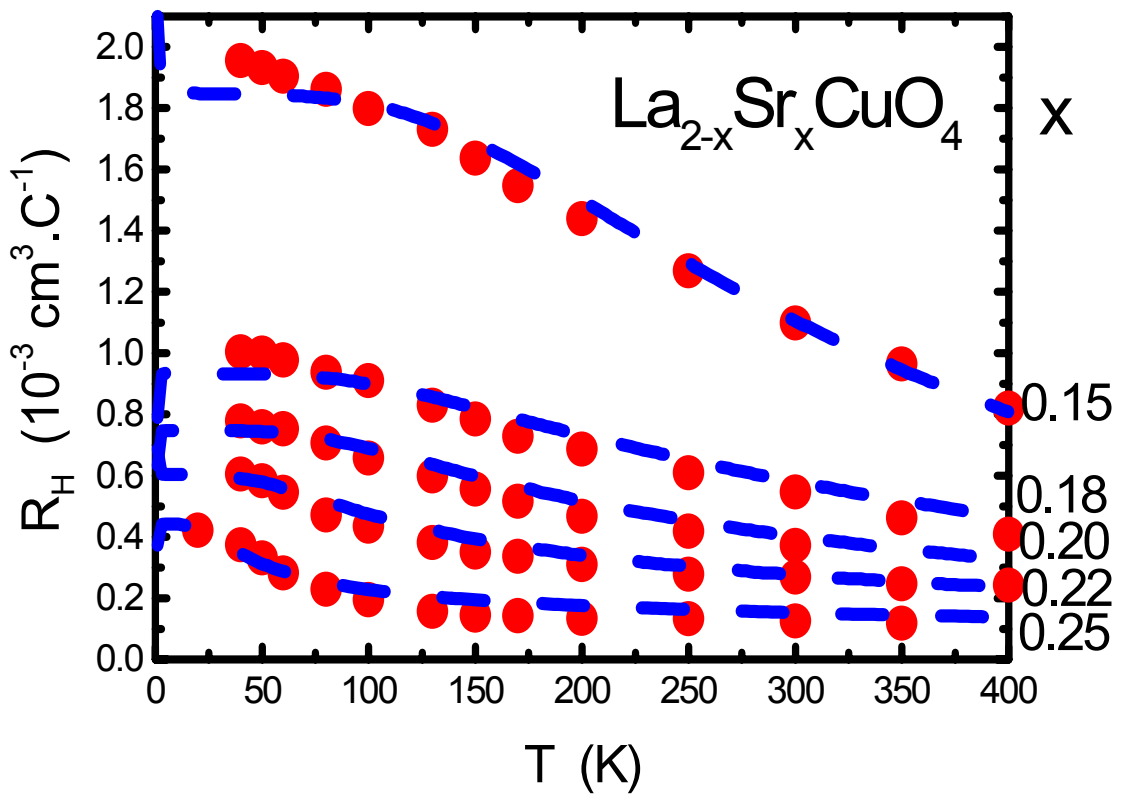

Figure 4 –

Filled circles: experimental $R_H(T)$ given by Hwang et al (Ref. 1), in polycrystalline $La_{2-x}Sr_xCuO_4$, for x= 0.15, 0.18, 0.20, 0.22, 0.25

Dashed lines: theoretical fits, the theoretical hole levels as the same as the experimental.

The calculations are made with: t = 0.23 eV, t' = 0.06 eV, 2t'/t = 0.52, $\Gamma(E_S) = 0.1$

## 4. Theoretical results and discussion

We use a theoretical band structure closed to the observed experimental one, but not in the fine details. We take a rigid band structure not varying with the doping, but we know that this variation occurs. Here we make our study with the ratio of transfer integrals of transfer closed to 2t'/t = 0.48 in order to obtain this special doping $ph_0 \approx 0.22$ when $E_F = E_S$ as in our previous studies, leading to convincing results [6].

In figures 1-4, we give the best fits with the parameters that we need for this. The value of $\Gamma$ maybe is too big because with our choice of t and t' the curvatures of the C.E.C. are not so pronounced as in reality.

But the aim of this paper is to demonstrate that the temperature dependence of the Hall coefficient is due to the effect of the distribution of the hole carriers in the electron-like energy levels and in the hole-like energy levels with increasing temperature. The results of our model do not change appreciably if we change slightly our set of parameters.

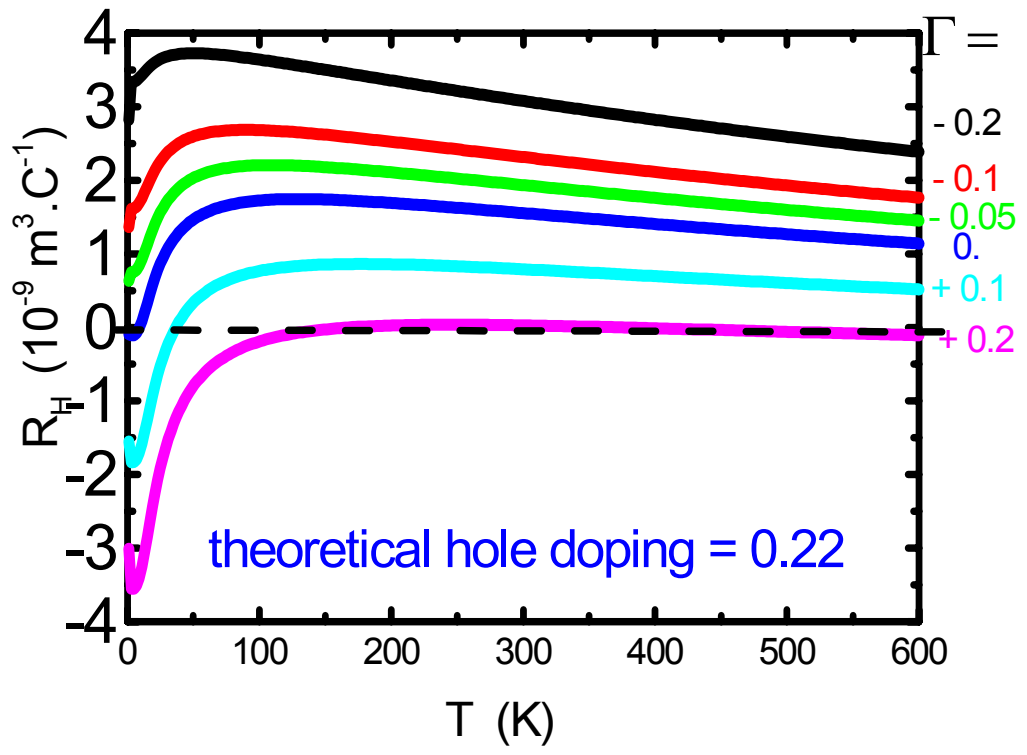

Figure 5 –

Theoretical curves of $R_H(T)$.showing the effect of Γ with , from the top to the bottom, the values: –0.2, -0.1, -0.05, 0, +0.1, +0.2. The calculations are made with: t = 0.23 eV, t' = 0.0553 eV, 2t'/t = 0.48, $V_{YBCO} \cong 174\ 10^{-30}\ m^3$, for a theoretical hole doping = 0.22.

Γ itself could change with the doping when the band structure varies. Near the optimum hole doped and overdoped systems Γ could decrease due to bigger curvatures of the C.E.C. In figure 5, we show the effect of the decreasing of Γ for a slightly overdoped system. This accounts for the behaviour of $R_H(T)$ in the optimum and slightly overdoped samples, where $R_H(T)$ is very flat and its value is very low closed to zero, and even can goes under zero at low or high temperature [1-3,5]. Theoretically this is due to the proximity of $E_F$ and $E_S$.

As the curvatures of the orbits increase, Γ goes from positive to negative value. This leads to very low (even negative) values to higher positive values at low T, for $R_H(T)$ in the optimum and overdoped samples.

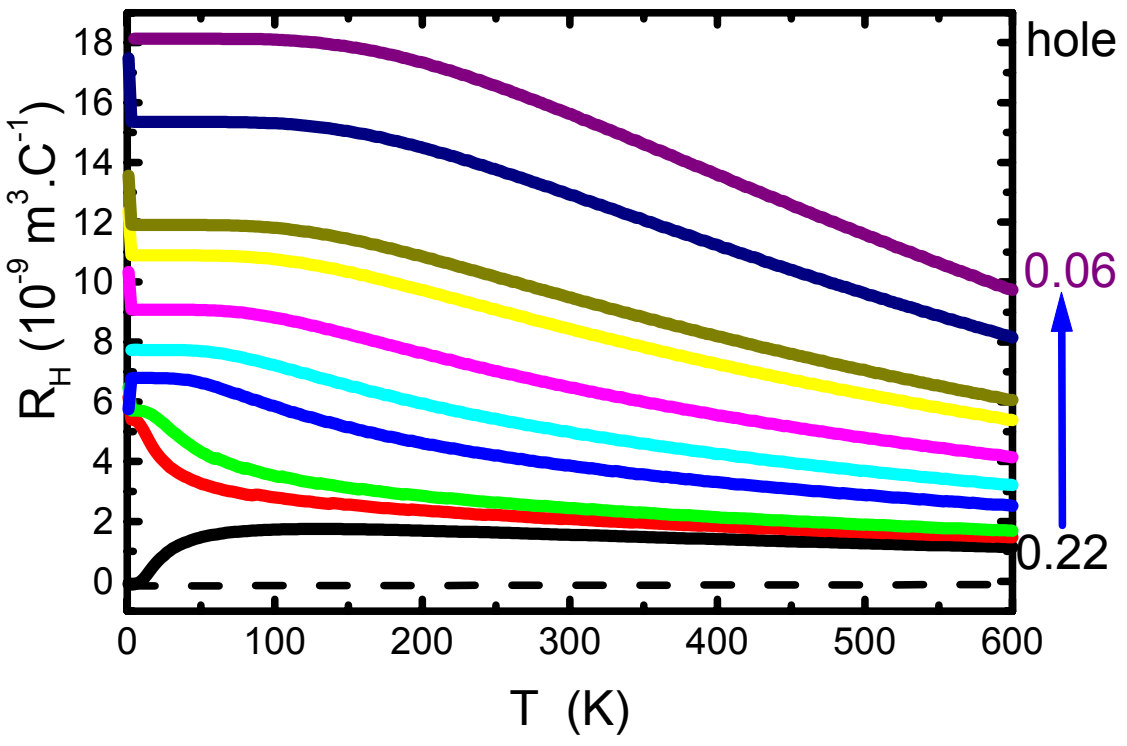

Figure 6 -

Theoretical curves of $R_H(T)$ with the same fit parameters of Figure 5 and with of $\Gamma(E_S) = 0$. From the bottom to the top the hole dopings are the following : 0.22, 0.20, 0.19, 0.16, 0.14, 0.12, 0.10, 0.09, 0.07, 0.06.

In figure 6, we show the theoretical $R_H(T)$ curves for a set of doping, using the same fit parameters as in Figure 5, letting $\Gamma(E_S) = 0$. We can see that the general behaviour of $R_H(T)$ is kept.

For very underdoped samples, near the metal-insulator transition our approach is no longer valid. We propose an explanation for the downturns observed in $R_H(T)$ [3,5] based on the localization of the carriers above an energy $E_{loc}$, due to the proximity of the metal-insulator transition (see Figure 7).

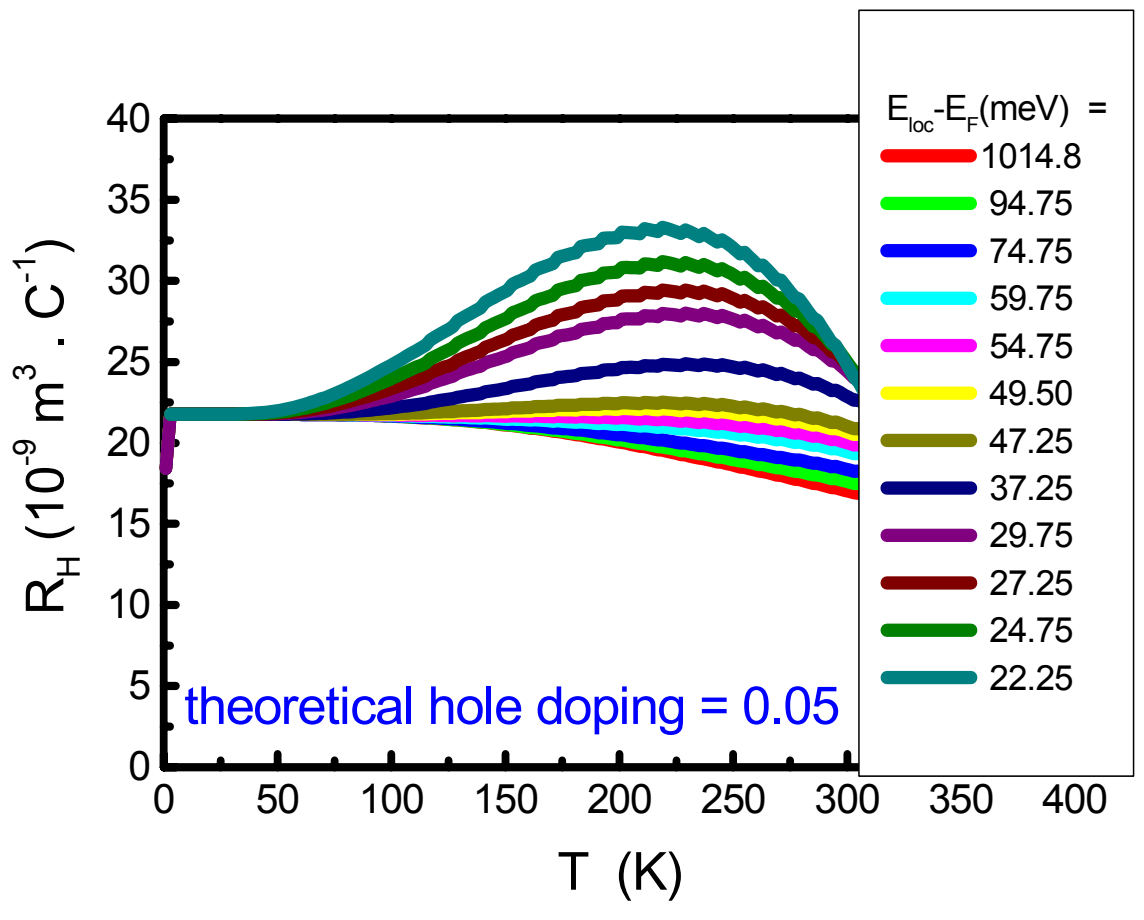

Figure 7

A localization level ($El_{oc}$) is introduced in the model to take into account the proximity of the metal-insulator transition in very underdoped sample.

The calculations are made for a hole doping of 0.05, with: t = 0.18 eV, t' = 0.04328 eV, 2t'/t = 0.48, $\Gamma(E_S) = 0$, $V_{YBCO} \cong 174\ 10^{-30}\ m^3$.

From the bottom to the top of the Figure 7 $E_{loc} - E_F$ varies from the infinity, that means no localization, to +22 meV, effective localization. We see that a strong maximum appears when the localization increases. This is due to the loss of localized particles, which do not contribute to transport.

## 5. Conclusion

In conclusion we find that the electronic structure of $CuO_2$ planes, with hole-like and electron-like orbits can explain the values of $R_H$ for the high Tc cuprates in the normal state and its temperature behaviour, this conclusion is reinforced by the fact that we obtain a representation of the experimental universal law $R_H(T)/R_H(T_0)$ versus $T/T_0$.